\documentclass[prd,twocolumn, tightten, aps, showpacs,preprintnumbers]{revtex4}

\newcommand{\be}{\begin{equation}}
\newcommand{\ee}{\end{equation}}
\newcommand{\bn}{\begin{eqnarray}}
\newcommand{\en}{\end{eqnarray}}

\newcommand{\ed}{\end{document}}
\newcommand{\nn}{\nonumber\\}

\begin{document}
\preprint{USM-TH-190}

\title{Lorentz Violating Julia-Toulouse Mechanism}
\author{Patricio Gaete}
\email {patricio.gaete@usm.cl} \affiliation{Departamento de
F\'{\i}sica, Universidad T\'ecnica F. Santa Mar\'{\i}a,
Valpara\'{\i}so, Chile}
\author{Clovis Wotzasek}
\email {clovis@if.ufrj.br} \affiliation{Instituto de F\'\i sica,
Universidade Federal do Rio de Janeiro, Brazil}
\date{\today}

\begin{abstract}
We propose a new Lorentz invariant violating extension for the
pure photonic sector of the Standard Model due to the condensation
of topological defects in the context of the Julia-Toulouse
mechanism. Possible physical consequences leading to direct
measurable effects over the confining properties of the elementary
particles are explored.
\end{abstract}
\maketitle


This work is devoted to study both the condensation of topological
defects(TD)\cite{Quevedo:1996uu} displaying Lorentz symmetry
violation (LIV)\cite{Kostelecky:1988zi} and the electromagnetic
properties of extended charged objects living inside the
condensate\cite{Gaete:2004dn}. It is shown that LIV--condensate
may result from a (modified) Julia-Toulouse mechanism (JTM) that
considers a non-homogeneous nucleation of TD. The condensate
characterizes the new vacuum of the theory now displaying LIV as a
spontaneous symmetry breaking (SSB) process with striking
consequences over the diferent phases of the electromagnetic
sector of an extended standard model. This breaking of Lorentz
symmetry (LS) can be parametrized by constant tensor fields
introducing the notion of a preferred frame - the rest frame of
the condensate\cite{Kostelecky:2001xz}. This mechanism of LIV
introduces the notion of {\it effective-brane} displaying a
jumping of dimension mechanism and allowing for the extended
charges to couple to tensor fields of lower ranks they normally
would in the presence of the Lorentz symmetry.

Lorentz invariance is one of the cornerstones of modern quantum
field theory (QFT). This seems to be an exact symmetry respected
by the Standard Model (SM) of elementary particles
\cite{Colladay:1996iz}. This theory has local LS and has been
confirmed to an extremely high precision. The observation of LIV
would therefore give a positive signature for the existence of a
new and unconventional physics. The necessity of a new scenario
has been suggested both to answer theoretical difficulties in
quantum gravity\cite{quantumgravity,astrobservation} and to the
exciting new possibility of high-precision experimental tests
searching for small violations of the LS \cite{Zavattini:2005tm}.
However, because LS is so deeply-rooted into the structure of QFT,
it is difficult to design a theory without a radical revision on
its foundations supporting those numerous experiments searching
for small LS violations arising as the low energy limit of such
yet unknown ultimate theory valid at the Planck scale
\cite{highenergy}.

Since our low-energy theories are relativistic QFT, it is
interesting to explore possible new phenomena in this framework
that could produce departures from exact Lorentz invariance. Among
these tentatives, probably the most studied framework is the
standard model extension (SME), consisting of the minimal SM plus
small Lorentz (and CPT) violating terms which has provided a
theoretical background for many experiments of Lorentz and CPT
violations such as: neutrino and neutral-meson oscillations,
clock-comparison tests on Earth and in space, the motion of a
spin-polarized torsion pendulum, hydrogen and antihydrogen
spectroscopy, QED in Penning traps,  cosmological birefringence,
microwave cavities, baryon asymmetry, muon properties,
etc\cite{testes}.

The basic idea in the SME is that the theory transforms normally
under rotations and boosts of the observer's inertial frame but
apparent Lorentz violations could appear when the particle fields
are rotated or boosted relative to the vacuum tensor expectation
values. This is called particle Lorentz transformations. One
interesting possibility is the SSB of LS in an ultimate
fundamental theory. Basically the interactions among tensor fields
trigger the formation of nonzero vacuum expectation values for
Lorentz tensors. The presence of these background quantities
throughout spacetime implies that Lorentz symmetry is
spontaneously broken.

The usual SM lacks the necessary dynamics to break LS
spontaneously, since the interactions triggering the Lorentz SSB
would destabile the empty vacuum. Spontaneous breaking could
however occur in more complicated theories involving extended
objects called p-branes known to have dynamics of the necessary
type. For most situations at energies well below the scale of the
underlying theory, it suffices to study the subset of the full
Standard-Model Extension for which the gauge structure of the
usual Standard Model are changed. Particularly interesting
examples are the LIV electrodynamics,  \cite{Carroll:1989vb} and
\cite{Colladay:1998fq}, including either even or odd
CPT--violating terms or both. These modified electrodynamics that
maintain the usual gauge invariance and are covariant under
observer Lorentz transformations read \bn {\cal L}_{\rm photon}
&=& - \frac 14 F_{\mu\nu}F^{\mu\nu} + \frac {\alpha}2 (b)^\kappa
\epsilon_{\kappa\lambda\mu\nu} A^\lambda F^{\mu\nu} \nonumber\\ &&
-\frac {\beta}4 (M)_{\kappa\lambda\mu\nu}
F^{\kappa\lambda}F^{\mu\nu} \label{pure_gauge}
\en
In \cite{Carroll:1989vb} a Maxwell-Chern-Simons like model ($\beta
= 0$) was studied to discuss relevant cosmological data while in
\cite{Colladay:1998fq} both odd and even CPT-violating terms were
included but emphasis was given to the $\alpha = 0$ case. These
theories have predicted several peculiar features that led to
sensitive tests of LIV.

The importance of the phenomenological model proposed here resides
on the fact that it provides us with an extension that comes from
a well-known physical mechanism very much associated with the dual
superconductivity. In this paper an extension for the
electromagnetic sector that breaks LS being parametrized by a
constante vector $v_\mu$ is proposed to control the dynamics of
the electromagnetic fields in the presence of LIV, considering the
vacuum as a Lorentz violating condensate.

We are then led to investigate possible mechanism such that a
reasonable quantum field theory might spontaneously break LI.
Taking inspiration from theoretical work on superconductivity as
the result of condensation of TD (dual Meisner effect) we argue
for the existence of theories with Lorentz invariant bare
Lagrangians where the formation of a condensate of TD, being
energetically favorable, leads to a non-Lorentz invariant vacuum
characterized by a constant vector induced by the condensation.
The new vacuum will display electric or magnetic properties
depending on the spacetime nature of this induced vector. The
final effective theory for the electromagnetic fields in this
LIV-vacuum reads,
\be
{\cal L}_{eff} = -\frac{1}{4} F_{\mu\nu}g^{\nu\beta} \left[
g^{\mu\alpha} + g^2\frac{v^{\mu}v^{\alpha} }{\Delta + m^2
}\right]F_{\alpha\beta} \label{effective} \ee This model is an
effective theory derived using an extension of the JTM in the
presence of broken symmetries.

The JTM is a condensation process dual to the Higgs mechanism,
proposed by Quevedo and Trugenberger (QT) \cite{Quevedo:1996uu}.
It is intended to describe phenomenologically the electromagnetic
behavior of anti-symmetric tensors in the presence of of
magnetic-branes (topological defects) that eventually condense due
to thermal and quantum fluctuations. Defects are classical
solutions of the equations of motion and may be classified as
topological or nontopological. Topological defects appear in
models that support spontaneous symmetry breaking. They are
important also in Cosmology and Condensed Matter Physics. In their
derivation of this mechanism, QT demanded a few features to take
into consideration the contributions of the fluctuations of the
TD. However, they did not worry about the exact (microscopic)
description of the condensation process. As so, the determination
of the critical values for the coupling constant or the density of
the condensate do not become an issue. Instead, the focus was in
the phenomenological description of the electromagnetic behavior
after brane condensation phenomenom had taken place. We shall
follow a similar atitude here. To study this nucleation process
the strategy is to look for a gauge invariant action, linear, with
up to two derivative dynamics that should recover the normal
vacuum in the limit of the dilute phase. Finally, among those
features, Lorentz symmetry was included before while here we shall
relax this last demand. Next, the consequences of this LIV
extension are examined. The LIV model studied here is obtained
following a previous work of the authors using the
JTM\cite{Gaete:2004dn} while the analytical work follows
\cite{GW2,Pato}.

In the JTM the topological defects play the role of magnetic
objects. They are characterized by their intensity, described by
the coupling constant $g$, and the Chern-Kernel $\Lambda_{q+1}$,
to be detailed below, describes their space dimension $p$ and
localization. In \cite{Quevedo:1996uu} it was found that, starting
from a Maxwell-like theory longitudinally coupled to p-branes, the
JTM produces a model for  free massive condensate:
\be
\int \left[F_{p+1}(A_p) - g \Lambda_{p+1}\right]^2 \to \int
\left[F_{p+2}^2(\Lambda_{p+1}) - {m^2} \Lambda_{p+1}^2\right], \ee
where $m$ is a free phenomenological mass parameter. Notice that
in the JTM, the Chern-Kernel $\Lambda_{p+1}$ is elevated to the
status of a propagating massive field, describing the low-energy
(long wavelength) hydrodynamical modes of the newly formed
condensate. The JTM is dual to the well known Higgs mechanism (HM)
where the gauge potential ``eats up" the degree of freedom of the
condensate. There is however a striking difference -- there is a
rank-jump here that is absent in the HM since $A_p
\to\Lambda_{p+1}$. In the JTM, on the other hand, by absorbing the
degrees of freedom of the original massless tensor $A_p$, it is
the condensate field that becomes massive. In this way the mass
density of the condensate becomes a free parameter to be
determined by the experiment. QT have shown that in the dilute
phase limit ($m\to 0$) one indeed recovers the original theory. In
summary, the JTM provides very successfuly phenomenological
description for the magnetic condensate through the effective
Chern-Kernel field dynamics.

In \cite{Gaete:2004dn} the present authors have used the formalism
proposed in \cite{Quevedo:1996uu} to study the interplay between
screening and confinement due to the condensation of topological
defects and try to fix the value of the free parameter $m$ by
connecting it to some observable of the theory: the measurement of
the confining effective potential of a pair of very massive test
charges. Using the QT phenomenology we were able to discuss the
dynamics of the extended charges ($p$-branes) inside the new
vacuum provided by the condensate. The model proposed in
\cite{Gaete:2004dn} considered the TD coupled both longitudinally
and transversally to two different tensor potentials, $A_p$ and
$B_q$, such that $p+q+2=D$ which, before condensation reads,
\footnote{For notational clarity we have dropped out all
normalizing factors and signs and adopted a simplified notation
proposed in \cite{Gaete:2004dn}.} \bn {\cal L}_d &\propto &
\left[F_{p+1}(A_p) - g \Lambda_{p+1}\right]^2\nn &+& g B_q
\epsilon^{q;1;p+1} \partial_1 \Lambda_{p+1} + H_{q+1}^2(B_q)\, ,
\en
where the current $J^q = \epsilon^{q;1;p+1} \partial_1
\Lambda_{p+1}$, represents the world-volume of a $(q-1)$-brane.
After condensation one obtains \bn {\cal L}_c &\propto &
H_{q+1}^2(B_q) + g B_q \epsilon^{q;1;p+1} \partial_1
\Lambda_{p+1}\nn &+& F_{p+2}^2(\Lambda_{p+1}) - {m^2}
\Lambda_{p+1}^2 \, ,
\en
showing an interacting theory displaying a $B\wedge F$-type of
coupling between the $B_q$ potential with the tensor
$\Lambda_{p+1}$ carrying the degrees of fredom of the condensate.
This theory displays a second mechanism of mass generation. It is
exactly the interplay between these two mass parameters that
controls the screening/confining character of the theory.

This is our starting point in order to study the nature of the
LIV--vacuum of the anti-symmetric tensor electromagnetic-like
theory. In \cite{Gaete:2004dn} the effective theory that results
from integrating out the fields representing the
vacuum-condensate, describing the dynamics of the quanta exchanged
by the test charges when immersed into the condensate, was given
as
\be
{\cal L}_{eff} \propto F_{q+1}(B_q) \left[1 + \frac {g^2}{\Delta +
m^2}\right] F^{q+1}(B_q)\, . \ee In this work we want to explore
the possibility that an analogous process occurs as the result of
an spontaneous breaking of the Lorentz symmetry. In their work, QT
assumed a homogeneous nucleation of the TD. We want to disccuss
now what would the outcome of condensation be in the presence of
symmetry violation. What is instrumental for the development of
SSB-JTM is the notion of ``effective brane". In the presence of a
constant rank-$m$ tensor, say $T_m$, a charge of dimension
$(q+m-1)$ will appear as an {\it effective brane} of dimension
$(q-1)$, $\Lambda_{p+1-m} \to \tilde\Lambda_{p+1} =
T_{[m}\Lambda_{p-m+1]}$. This lower-dimensional brane represents
the intersection of the subspace transverse to the $T_m$ tensor
with the world-volume of the $(q+m-1)$-brane giving the
world-volume for the effective $(q-1)$-brane. Then, for instance,
for a constant vector, $T_1 \equiv v_1$, the world-volume of a
membrane is changed to an effective string, that now couples to
Kalb-Ramond potentials and a string behaves as if it were a
particle, coupling to vector fields and so on. The obvious
consequence of this mechanism is that higher-dimensional objects
are now able to couple to lower-ranked potentials. Therefore, upon
nucleation, this dimension-changing mechanism for the branes
becomes a rank-changing mechanism for the condensates. This
rank-changing mechanism alters the JTM that now has a condensed
action given as \bn {\cal L}_{SSB} &\propto & H_{q+1}^2(B_q) + g
T_{[l}B_{q]} \epsilon^{q;1;p+1} \partial_1 \Lambda_{p-l+1}\nn &+&
F_{p-l+2}^2(\Lambda_{p-l+1}) + {m^2} \Lambda_{p-l+1}^2
\en
For a condensate, where LIV is characterized by a constant vector
$v_1 \equiv v_\mu$ we obtain, \bn {\cal L}_{LIV} &=&
H_{q+1}^2(B_q) + g v_{[1}B_{q]} \epsilon^{q+1;1;p} \partial_1
\Lambda_{p}\nn &+& F_{p+1}^2(\Lambda_{p}) + {m^2} \Lambda_{p}^2 \,
.
\en
Upon integration of the field charaterising the LIV-condensate we
obtain the result expressed before in (\ref{effective}) Next we
shall study the consequences of this mechanism over the confining
and screening phases of the charges living inside this condensate.
To this end we shall focus on $D=3$ space-time models that might
be of revelevance in condensed matter physics. We now examine the
interaction energy between static pointlike sources for the model
under consideration, along the lines of Refs.\cite{GW2}. This can
be done by computing the expectation value of the energy operator
$H$ in the physical state $\left| \Phi \right\rangle$ describing
the sources, which we will denote by $ \left\langle H
\right\rangle_\Phi$.

The starting point is the 3D JTM Lagrangian density:
\begin{equation}
{\cal L} =  - \frac{1}{4}F_{\mu \nu }^2  + \frac{g}{2}\phi
\varepsilon ^{\mu \nu \lambda } v_\mu  F_{\nu \lambda }  +
\frac{1}{2}\partial _\mu  \phi \partial ^\mu  \phi  - \frac{{m^2
}}{2}\phi ^2, \label{LVJ10}
\end{equation}
where $m$ is the mass for the condensate field $\phi$. We restrict
ourselves to static scalar fields, a consequence of this is that
one may replace $\Delta \phi = - \nabla ^2\phi$, with $\Delta
\equiv\partial _\mu
\partial ^\mu$. It also implies that, after performing the integration
over $\phi$, the effective Lagrangian density is
given by
\begin{equation}
{\cal L}_{eff} =  - \frac{1}{4}F_{\mu \nu }^2  - \frac{{g^2 }}{8}V^{\nu
\lambda } F_{\nu \lambda } \frac{1}{{\nabla ^2  - m^2 }}V^{\gamma
\beta } F_{\gamma \beta } + A_\mu J^\mu, \label{LVJ15}
\end{equation}
where $V^{\nu \lambda}  \equiv \varepsilon ^{\mu \nu\lambda } v_\mu $ and $J^\mu$ is the external current of the test charges and $(\mu ,\nu  = 0,1,2)$ and
$(i,k= 1,2)$.
It should be clear that the theory (\ref{LVJ15}) depends on the spacetime character of the constant vector $v_\mu$. We shall consider different possibilities next.

{\underline {\bf Spacelike background case}}. To obtain the interaction energy in the $V^{0i} \ne 0$ and $V^{ij}=0$ case, we compute the Hamiltonian density of (\ref{LVJ15})
\begin{equation}
\begin{array}{r}
 {\cal H} = \left\{ {\frac{1}{2}\Pi ^2  + \frac{{g^2 }}{2}\frac{{\left( {{\bf V} \cdot {\bf \Pi} } \right)^2 }}{{\left( {\nabla ^2  - M^2 } \right)}} + \frac{1}{2}B^2 + c(x)\Gamma(x)} \right\}
 \end{array}, \label{LVJ25}
\end{equation}
where $\Gamma _1 \left( x \right)
\equiv\partial _i \Pi ^i - J^0 \approx 0$ is the Gauss law, $M^2\equiv m^2  + g^2 V^2$ and $B$ is the magnetic field. 
To fix gauge symmetry we adopt the linear gauge discussed in our previous works \cite{Pato,Gaete:2004dn}.
Next the physical state is constructed, following Dirac \cite{Dirac2},
\begin{equation}
\begin{array}{l}
\left| \Phi  \right\rangle  \equiv \left| {\overline \Psi  \left(
\bf y \right)\Psi \left( {\bf y}\prime \right)} \right\rangle  \\
 = \overline \psi \left( \bf y \right)\exp \left(
{ie\int\limits_{{\bf y}\prime}^{\bf y} {dz^i } A_i \left( z
\right)} \right)\psi \left({\bf y}\prime \right)\left| 0
\right\rangle, \label{LVJ40}
 \end{array}
 \end{equation}
where $\left| 0 \right\rangle$ is the physical vacuum state and
the line integral appearing in the above expression is along a
spacelike path starting at ${\bf y}\prime$ and ending at $\bf y$,
on a fixed time slice.

We now turn to the problem of obtaining the interaction energy
between pointlike sources in the model under consideration. A fermion
is localized at ${\bf y}\prime$ and an antifermion at ${\bf y}$.
One might show that the interaction energy is
\bn
\left\langle H \right\rangle _\Phi   = \left\langle \Phi
\right|\int {d^2 x} \left\{ {\frac{1}{2}{\bf \Pi} ^2  + \frac{{g^2
}}{{2}}\frac{{\left( {{\bf V} \cdot {\bf \Pi} } \right)^2 }}{{
(\nabla ^2 - M^2) }} + \frac{1}{2} B^2 } \right\}\left| \Phi
\right\rangle\nn
\label{LVJ50}
\en
where $\left\langle H
\right\rangle _\Phi =  \left\langle H
\right\rangle _0  + V^{\left( 1 \right)}  + V^{\left( 2 \right)}$  with $\left\langle H \right\rangle _0  = \left\langle 0
\right|H\left| 0 \right\rangle$ giving the background vacuum term while the contributions of the test charges are given by
$V^{\left( 1 \right)}$ and $V^{\left( 2 \right)}$ and read,
\begin{equation}
V^{\left( 1 \right)}  = - \frac{e^2}{{2\pi }}K_0 \left( {M|{\bf y}
- {\bf y}^{\prime}  |} \right). \label{LVJ65}
\end{equation}
and
\begin{equation}
V^{\left( 2 \right)}  = \frac{{e^2 m^2 }}{{4M }}|{\bf y} - {{\bf
y}^\prime}|. \label{LVJ70}
\end{equation}
By putting together Eqs.(\ref{LVJ65}) and (\ref{LVJ70}), we obtain
for the total interaction energy  ($L\equiv|{\bf y}-{\bf {y^\prime}}|$)
\begin{equation}
V(L) = - \frac{{e^2 }}{{2\pi }}K_0 \left( {ML} \right) +
\frac{{e^2 m^2 }}{{4M}}L \label{LVJ75}\, .
\end{equation}

{\underline {\bf Timelike background case}}. Now we focus on the case $V^{0i}=0$ and $V^{ij}\ne 0$. The corresponding Lagrangian density reads
\begin{equation}
{\cal L} =  - \frac{1}{4}F_{\mu \nu } F^{\mu \nu }  - \frac{{g^2
}}{{8}}V^{ij} F_{ij} \frac{1}{{ \nabla ^2  - m^2 }}V^{kl} F_{kl} -
A_0 J^0, \label{LVJ80}\nonumber
\end{equation}
Following the same steps that led to (\ref{LVJ75}) we find that the
potential for two opposite charges located at $\bf y$ and $\bf y\prime$ reads ($\mu$ is a massive cutoff regulator)
\begin{equation}
V = \frac{{e^2 }}{{2\pi }}\ln(\mu L). \label{LVJ100}
\end{equation}

Therefore, LIV models with constant spacelike-vectors lead to
linear confinement of the test charges while the spacelike case
remains coulombic. It is important to mention that a similar study
made in $D=4$ models led to analogous results\cite{breve}.

In summary, in this work we constructed a low energy effective
theory that displays Lorentz invariance breaking as the result of
condensation of topological defects and studied its physical
consequences. The resulting theory exhibits a new extension for
the electromagnetic sector of the standard model that is
alternative to the extensions proposed so far. This construction
uses a condensation mechanism similar to the JTM and describes how
a stable effective theory may lead to spontaneous breaking of
Lorentz invariance due to the condensation phenomenon, being
characterized by constant tensors of any rank. A physical
observable related to the existence of this new vacuum was studied
which may lead to possible measurable predictions of the Lorentz
breaking. As a particular application we have considered the
confinement and screening issue for a three-dimensional vector
gauge that couples to strings through this LIV-condensation
process. The condensation of magnetic strings leads to a massive
scalar condensate via this modified JTM. The breaking of Lorentz
symmetry is manifest through a constant vector, giving a preferred
spacetime direction with distinct consequences. We found that when
the condensate constant vector $v_\mu$ is purely timelike the
resulting static potential remains Coulombic while it was found
that when it is purely spacelike the static potential displays a
linear confining piece. This result survives any dimensional
consideration being valid also in our four-dimensional world.
However, in the four-dimensional case we found that the
condensation of magnetic string gives rise to a massive vectorial
condensate. Since we do not see confined charges, the assumption
that Lorentz symmetry is being broken by a constant vector leads
us to support\cite{breve} a timelike type of vector-breaking as
suggested in \cite{Carroll:1989vb}.

This work is partially supported by Fondecyt (Chile) grants
1050546 and 7060050. C.W. would like to thank CNPq/PRONEX/FAPESQ
for financial support.

\end{document}